\documentclass[12pt]{article}
\pdfoutput=1

\usepackage{amsmath,amsfonts,amssymb,amsthm,amssymb,bbm,bm}
\usepackage{pdfsync}
\usepackage{graphicx,import}
\usepackage[latin1]{inputenc}
\usepackage{hyperref}
\usepackage[all]{xy}
\usepackage{stmaryrd}
\usepackage{rotating}
\usepackage{color}  
\usepackage{slashed}
\usepackage{cancel}
\usepackage{array, makecell} %
\usepackage{comment}
\usepackage{tikz-cd}
\usepackage{hhline}

\newcommand{\p}{\partial}

\newcommand{\bit}{\begin{itemize}}
\newcommand{\eit}{\end{itemize}}
\newcommand{\bd}{\begin{description}}
\newcommand{\ed}{\end{description}}

\newcommand{\bc}{\begin{center}}
\newcommand{\ec}{\end{center}}

\newcommand{\scr}{\scriptstyle}

\newcommand{\be}{\begin{equation}}
\newcommand{\ee}{\end{equation}}
\newcommand{\bea}{\begin{eqnarray}}
\newcommand{\eea}{\end{eqnarray}}
\newcommand{\bs}{\begin{subequations}}
\newcommand{\es}{\end{subequations}}
\newcommand{\nn}{\nonumber}

\newcommand{\w}{\wedge}

\newcommand{\f}{\frac}
\newcommand{\tl}{\tilde}
\def\p{\partial}

\newcommand{\na}{\nabla}

\newcommand{\sd}{\slashed{\delta}}

\def\a{\alpha}
\def\b{\beta}
\def\g{\gamma}
\def\d{\delta}
\def\eps{\epsilon}

\def\th{\theta}

\def\k{\kappa}
\def\l{\lambda}
\def\m{\mu}
\def\n{\nu}

\def\r{\rho}

\def\s{\sigma}

\def\om{\omega}
\def\G{\Gamma}

\def\Si{\Sigma}

\def\Om{\Omega}

\newcommand{\scri}{\cal I}

\newcommand{\FE}[1]{\mathbbm E^{(#1)}}

\newcommand{\standard}{standard }

\begin{document}

\title{\Large{\bf A note on dual gravitational charges}}

\author{\large{Roberto Oliveri$^{a}$ and Simone Speziale$^{b}$}
\smallskip \\ \small{$^a$CEICO, Institute of Physics of the Czech Academy of Sciences, Prague, Czech Republic } \\
\small{$^b$Aix Marseille Univ., Univ. de Toulon, CNRS, CPT, Marseille, France} 
}

\maketitle

\begin{abstract}

Dual gravitational charges have been recently computed from the Holst term in tetrad variables using covariant phase space methods.
We highlight that they originate from an exact 3-form in the tetrad symplectic potential that has no analogue in metric variables. 
Hence there exists a choice of the tetrad symplectic potential that sets the dual charges to zero. This observation relies on the ambiguity of the covariant phase space methods.
To shed more light on the dual contributions, we use the Kosmann variation to compute (quasi-local) Hamiltonian charges for arbitrary diffeomorphisms. 
We obtain a formula that illustrates comprehensively why
the dual contribution to the Hamiltonian charges: (i) vanishes for exact isometries and asymptotic symmetries at spatial infinity;
(ii) persists for asymptotic symmetries at future null infinity, in addition to the usual BMS contribution. 
Finally, we point out that dual gravitational charges can be equally derived using the Barnich-Brandt prescription based on cohomological methods, and that the same considerations on asymptotic symmetries apply.

\end{abstract}

\tableofcontents

\section{Introduction}

Recent work \cite{Godazgar:2020gqd,Godazgar:2020kqd} has shown that, if one starts from the first-order tetrad Lagrangian including a dual term (often called Holst term), the Hamiltonian charges at future null infinity contain contributions from both the standard BMS terms and the new dual gravitational charges defined in \cite{Godazgar:2018qpq}. These were further studied in \cite{Godazgar:2018vmm,Kol:2019nkc,Godazgar:2019dkh}. For previous related work, see e.g. \cite{ashtekar1982nut,Barnich:2008ts,Corichi:2013zza,Frodden:2017qwh}.
The calculation in \cite{Godazgar:2020gqd,Godazgar:2020kqd} uses covariant phase space methods \cite{kijowski1976,Ashtekar:1981bq,Iyer:1994ys}.
These dual contributions to the charges, or dual charges for short, are present only in tetrad variables, and not in metric variables.
The first goal of this brief note is to highlight that the origin of this difference lies in the more general discrepancy between the metric and tetrad symplectic potentials, which differ by a certain exact 3-form. In covariant phase space methods one is free to add exact 3-forms to the symplectic potential, therefore there exists a choice for tetrad variables that reproduces all results of the metric phase space, with vanishing dual charges. This observation exposes an ambiguity of the dual contribution to Hamiltonian charges.

The difference in symplectic potentials is partially tamed by the fact that isometries in tetrad variables are not given by a simple Lie dragging, but require also fixing the internal Lorentz transformations. The combined transformation is sometimes referred to as Kosmann derivative in the literature;
see e.g.~\cite{TedMohd,Prabhu:2015vua,Hajian:2015xlp,Barnich:2016rwk}, where it was shown that this prescription correctly reproduces the metric Noether charges for any diffeomorphism, and the metric Hamiltonian charges for isometries
 (see also \cite{Frodden:2017qwh,Aneesh:2020fcr}).
For asymptotic symmetries at spatial infinity, the  metric Hamiltonian charges (namely the Poincar\'e charges) are reproduced using either the Kosmann or the standard Lie derivative variations: their difference vanishes in the limit \cite{Ashtekar:2008jw,Corichi:2013zza,DePaoli:2018erh}. 
For asymptotic symmetries at future null infinity, the Kosmann and Lie derivative variations again coincide thanks to the fall-off conditions, which set to zero the internal charges. However, the metric Hamiltonian charges (namely the BMS charges) are not exactly reproduced, because there remains the additional dual contribution \cite{Godazgar:2020kqd}.

Our second goal is to better understand this state of affairs: why shifting from the Lie transformation to the Kosmann transformation is needed for exact isometries but not for asymptotic symmetries, and why the dual contribution survives at null infinity but not at spatial infinity.
To that end, we compute the Hamiltonian charges associated to the Kosmann transformation for an arbitrary diffeomorphism and on an arbitrary 2d surface. We obtain a formula that allows one to understand all different behaviours at a glance. 
In particular, we show that for arbitrary diffeomorphisms, the Kosmann prescription for Hamiltonian charges in tetrad variables fails to reproduce the metric expressions in both standard and dual contributions. Explicit use of the fall-off 
and gauge-fixing
conditions restores the metric result for both standard and (vanishing) dual contributions at spatial infinity; and only for the standard contribution at future null infinity, leaving a non-vanishing dual contribution.

To complete our analysis, we consider  the cohomological methods used to define charges \`a la Barnich-Brandt (BB) \cite{anderson1992introduction,Anderson:1996sc,Barnich:2000zw,Barnich:2001jy}. This addresses a question that was left open in \cite{Godazgar:2020kqd}.
We point out that the general expression for the BB charges for the first-order tetrad Lagrangian with the Holst term is identical to the one for the Hamiltonian charges obtained with covariant phase space methods. Therefore the same considerations highlighted above apply: using the Kosmann variation, the charges associated with isometries and asymptotic symmetries at spatial infinity match the metric ones, with no dual contributions, whereas the fall-off conditions of \cite{Godazgar:2020kqd} at future null infinity lead to a dual contribution to the BB charges in addition to the standard BMS one.

The existence of a dual contribution to the BMS charges using cohomological methods allows one to argue against their ambiguity. 
There is in fact in this case a preferred choice to avoid the ambiguities associated with freely adding exact 3-forms to the symplectic potential (picking the weakly-vanishing Noether current). Standing by this choice, it is not possible to set the dual contribution to zero, like it was possible with covariant phase methods alone. 
However, we point out that the dual contribution vanishes exactly if one uses cohomological methods with a second-order tetrad Lagrangian. This might come as a surprise, because it has been proved that the charges obtained from cohomological methods for exact symmetries do not depend on the order of the Lagrangian chosen \cite{Barnich:2000zw}. Our analysis shows that the situation is different for the case of asymptotic symmetries: the presence of a dual contribution depends on the order of the Lagrangian. 
On the other hand, the standard metric BMS contribution is obtained from  both first and second order Lagrangians. Therefore the dual contribution  appears to be more fragile than the standard contribution, and heavily depending of choices which are equivalent from the point of view of field equations alone.

To highlight our main messages, we restrict attention to General Relativity without torsion extensions. Including torsion can be easily done following the formulas provided in \cite{Oliveri:2019gvm}, and does not change our main message. The presence of torsion introduces a non-vanishing dual contribution to the Hamiltonian charges already in metric variables. But the difference of the symplectic potentials by an exact 3-form remains, therefore this dual contribution is inequivalent to the one computed using tetrad variables.
For the related topic of the contribution of torsion to the first law of black hole mechanics, see \cite{DeLorenzo:2018odq,Aneesh:2020fcr}.

{\bf{Note added in v2.}} In a first version of this paper available
on-line, we left open the possibility of differences between the
metric and tetrad standard contributions to the BMS charges at
subleading orders. These differences are excluded thanks to
non-trivial properties satisfied by the BMS vector fields at all
orders. We thank Mahdi Godazgar for discussions of this point.

\section{Tetrad and metric covariant phase spaces}

\subsection{Brief review of first-order formalism}
The tetrad and metric Lagrangians in the first-order formalism have independent connection variables, and are given, respectively, by
\begin{align}\label{Le}
L^{\scr (e, \g)} & :=L^{\scr e}+\f1\g\tl L^{\scr e}
= \f12 \epsilon_{IJKL}~e^I\w e^J\w F^{KL}(\om) + \f1\gamma ~e^I\w e^J\w F_{IJ}(\om), \\ \label{Lg}
L^{(g,\g)} &:=L^{\scr g}+\f1\g\tl L^{\scr g}
=g^{\m\n}R_{\m\n}(\G) \eps - \f1{2\g} \eps^{\m\n\r\s}R_{\m\n\r\s}(\G) \eps.
\end{align}
The term proportional to $1/\gamma$ is often called Holst term in the literature \cite{Holst}, and we use the conventions of \cite{Oliveri:2019gvm} with $\eps$ the volume 4-form. We use a tilde to distinguish the dual Lagrangian from the \standard one.
A similar notation will be used for the potentials and charges.

Taking arbitrary variations of the Lagrangian over the independent variables, one gets the field equations and a boundary term $d\th(\d)$.
The latter is used to read the symplectic potential $\th(\d)$ in covariant phase phase methods. More precisely, the symplectic potential is the integral of $\th(\d)$ on the hypersurface $\Si$ used to describe the phase space, but we will use the same term for $\th(\d)$ as well.
The Lagrangians \eqref{Le} and \eqref{Lg} define in this way two different phase spaces, with potentials given respectively by 
\begin{subequations}\label{the}\begin{align}
& \th^{(e,\g)}(\d)= \th^e+\f1\g\tl\th^e = P_{IJKL}~e^I\w e^J\w \d\om^{KL}, \\
& \th^{(e,\g)\m}(\d)=-\f16 \eps^{\m\n\r\s} \th^{(e,\g)}_{\n\r\s} = 2e^{[\m}_I e^{\n]}_J\d\om_\n^{IJ} -\f1\g \eps^{\m\n\r\s}e_{\n I}e_{\r J}\d\om_\s^{IJ},
\end{align}\end{subequations}
where $P_{IJKL} = \f12\eps_{IJKL} + \f1\g\eta_{I[K}\eta_{L]J}$, and
\begin{subequations}\label{thg}\begin{align}
& \th^{(g,\g)}_{\n\r\s}(\d)= \th^g+\f1\g\tl\th^g = \th^{(g,\g)\m}(\d) \eps_{\m\n\r\s}, \\
& \th^{(g,\g)\m}(\d) = \left(2g^{\r[\m}g^{\n]\s} -\f1\g \eps^{\m\n\r\s}\right) g_{\r\l} \d \G^\l_{\n\s}.
\end{align}\end{subequations}

By construction, the Lagrangian prescribes the symplectic potential up to the addition of an exact 3-form. To stress this fact, we refer to the choices \eqref{the} and \eqref{thg} as `bare' choices. 
The bare choices for two equivalent Lagrangians like \eqref{Le} and \eqref{Lg}, with the same set of physical solutions, need not be equal, and in fact they turn out not to be. Using the defining relations between the tetrad and the metric, and the corresponding one for the connections,  \eqref{the} and \eqref{thg} differ by an exact 3-form  \cite{DePaoli:2018erh,Oliveri:2019gvm},
\begin{align}\label{thethg}
\th^{(g,\g)}(\d) &=\th^{(e,\g)}(\d)+d\a(\d),\\
\label{alfa}
\a(\d) &= -P_{IJKL}~e^I\w e^J \left(e^{\r K}\d e^L_\r\right) =\star (e_{I}\w \d e^{I}) +\f1\g e_{I}\w \d e^{I} :=\a_\star(\d) +\f1\g\a_\g(\d).
\end{align}
This relation\footnote{Here written in the absence of torsion. If torsion is present, there is an additional bulk term; see Eq.~(VIII.14) of \cite{Oliveri:2019gvm}.}
has important consequences for covariant phase space methods, because two symplectic potentials that differ by an exact 3-form define inequivalent symplectic structures,
\be \label{Om}
\Om^{(g,\g)}(\d_1,\d_2) = \Om^{(e,\g)}(\d_1,\d_2)+\Om^\a(\d_1,\d_2)\neq  \Om^{(e,\g)}(\d_1,\d_2).
\ee
The canonical generator of a gauge transformation $\d_\eps$ is defined via
$\sd H_{\eps}:=\Om(\d,\d_\eps)$. Therefore Eq.~\eqref{Om} implies that the tetrad and metric phase spaces carry different Hamiltonian charges a priori.
The difference has been studied already in the literature \cite{TedMohd,Prabhu:2015vua,DePaoli:2018erh,Oliveri:2019gvm}, and we report here only the formulas we need for our present discussion. 
We will first consider arbitrary gauge transformations and finite regions, and afterwards focus the discussion on isometries and asymptotic symmetries. Also, we will use the term charge in a rather loose sense, referring to the defining equation $\sd H_{\eps}:=\Om(\d,\d_\eps)$. 
Our considerations and main message concern the general set-up preceding the specific integrability issues. If need be, one can always restrict attention to diffeomorphisms tangent to the 2d surface, for which integrability is immediate.

Before moving on, let us remark that the difference in bare symplectic potentials shows up only because we are using a first order formalism. 
To compare the situation with the second order formalism, we write $\om^{IJ}=\om^{IJ}(e)+C^{IJ}$, where $\om^{IJ}(e)$ is the Levi-Civita connection and $C^{IJ}$ is the contorsion. 
Using $F^{IJ}(\omega) = F^{IJ}(e)+ d_{\omega(e)}C^{IJ} + C^{I K}\w C_{K}{}^{J}$, and changing variables from $\om^{IJ}$ to $C^{IJ}$, the tetrad Lagrangian \eqref{Le} becomes 
\begin{align}\label{LeC}
L^{(e,C,\g)} &= P_{IJKL} \Si^{IJ}\w \left(F(e)^{KL}+C^{KM}\w C_M{}^L\right) + d\left(P_{IJKL} \Si^{IJ}\w C^{KL}\right)\nn\\
&= L^g +P_{IJKL} \Si^{IJ} C^{KM}\w C_M{}^L + d\left(P_{IJKL} \Si^{IJ}\w C^{KL}\right),
\end{align}
with the shorthand notation $\Sigma^{IJ} := e^I \w e^J$.
The bare symplectic potential is
\be
\th^{(e,C,\g)}(\d) = \th^{g}(\d) + \d\left(P_{IJKL}\Si^{IJ} \w C^{KL}\right).
\ee
Since the second term is a total variation in field space, it does not affect the symplectic structure, which is then equivalent to the metric one.
We conclude that the second-order tetrad Lagrangian will always produce the same charges as the metric theory
with vanishing torsion.

\subsection{The origin of dual charges} 
\label{dualorigin}
\paragraph{Metric charges}
The metric Lagrangian \eqref{Lg} is invariant under diffeomorphisms given by the ordinary Lie derivative, $\d_\xi=\pounds_\xi$. 
Using the metric symplectic potential \eqref{thg}, the on-shell result with vanishing torsion is  \cite{Iyer:1994ys}
\begin{align}
\sd H_{\xi}^{(g,\g)} &= \Om^{(g,\g)}(\d,\d_\xi)
= \int_{\p\Si} \d\k_\xi -i_\xi \th^{(g,\g)}(\d)  \nn \\\nn
&= - \int_{\p\Si} \f12\eps_{\m\n\r\s}\big[ (\d\ln{\sqrt{-g}}) \na^\r\xi^\s + \d g^{\r\a}\na_\a\xi^\s 
\\ &\qquad\qquad\qquad\quad + \xi^\r \left(\na_\a \d g^{\a\s} + 2 \na^\s \d\ln\sqrt{-g}\right)  - \xi_\a\na^\r\d g^{\s\a} \big] dx^\m\w dx^\n \label{Hg}\\
&=\sd H_\xi^g,\nn
\end{align}
where 
$\k_\xi = -\f12 \eps_{\m\n\r\s} \na^\r\xi^\s dx^\m\w dx^\n = - \f12 \eps_{IJKL} \Si^{IJ} D^K\xi^L  $
is the Komar 2-form, which also coincides with Wald's definition of the (metric) Noether charge for diffeomorphisms.
We observe that there are no dual metric charges, nor dependence of $\sd H_{\xi}^{(g,\g)}$ on $\g$, because the $\g$-term of the metric symplectic potential \eqref{thg} is proportional to the torsional part of the connection, $T^\m{}_{\n\s} = 2\G^\m_{[\n\s]}$, and therefore vanishes on-shell.
This is why dual gravitational charges are not present in the metric formulation without torsion,
as pointed out in \cite{Godazgar:2020gqd}. 

\paragraph{Tetrad charges}
The tetrad Lagrangian \eqref{Le} is invariant under diffeomorphisms given by the ordinary Lie derivative, $\d_\xi=\pounds_\xi$, and also under internal Lorentz transformations, which we denote by $\d_\l$. It is at times convenient to 
take a linear combination of the two transformations, which results in the covariant Lie derivative
$L_\xi := \pounds_\xi + \d_{i_\xi\om}.$
Using the tetrad potential \eqref{the}, 
the Hamiltonian charges for these gauge symmetries 
satisfy
\begin{align}
\sd H_{(\l,\xi)}^{(\scr e,\g)} &= \Om^{(e,\g)}(\d,\d_{\l}+\d_{\xi}) = \int_{\p\Si} \d q^{(e,\g)}_\xi -i_\xi \th^{(e,\g)}(\d) +\d q^{(e,\g)}_\l \nn \\
&= \int_{\p\Si} \f12\eps_{IJKL}  \left[(i_\xi\om^{IJ}-\l^{IJ}) \d\Si^{KL}  - i_\xi \Si^{IJ}\w \d\om^{KL}\right]  \nn \\  \label{He}  
&\quad + \f1\g \int_{\p\Si} (i_\xi\om^{IJ} -\l^{IJ})\d\Si_{IJ}- i_\xi \Si_{IJ}\w \d\om^{IJ} \\
&= \sd H_{(\l,\xi)}^e+ \f1\g\sd \widetilde H_{(\l,\xi)}^e, \nn
\end{align}
where $q^{(e,\g)}_\xi=P_{IJKL}\, i_\xi\om^{IJ}\Si^{KL}
=q^{e}_\xi+ \tl q^{e}_\xi / \g $ and
$q^{(e,\g)}_\l=P_{IJKL}\,\l^{IJ}\Si^{KL}$
are the tetrad Noether charges for diffeomorphisms and internal Lorentz transformations.

The generic inequivalence of the charges defined by Eq.~\eqref{Hg} and Eq.~\eqref{He} should be manifest.
First of all, Eq.~\eqref{Hg} has no internal Lorentz charges, as this gauge symmetry is absent in the metric formalism. 
Second, the standard charges are different, for instance Eq.~\eqref{Hg} contains derivatives of $\xi$, while Eq.~\eqref{He} does not.
The difference can be further highlighted manipulating the terms in Eq.~\eqref{He} to obtain 
\be\label{HeS}
\sd H_{\xi}^e = \int_{\p\Si} \f12\eps_{IJKL}  \left(i_\xi\om^{IJ} \d\Si^{KL}  - i_\xi \Si^{IJ}\w \d\om^{KL}\right)
=\int_{\p\Si}\d q^e_\xi -i_\xi \th^e(\d).
\ee
We see that both the Noether charge and the potential appearing here are different from the metric ones, a difference not removed by their combination. 
Nonetheless, there are circumstances in which Eq.~\eqref{HeS} reproduces Eq.~\eqref{Hg}, most notably asymptotic symmetries at spatial \cite{Ashtekar:2008jw} and null \cite{Godazgar:2020kqd} infinity. 
More on this in subsection~\ref{AS}.

Finally and more importantly for our discussion, non-trivial dual 
gravitational charges are now present, 
\be\label{dualC}
\sd \widetilde H^e_{\xi} = \int_{\p\Si}  \left(i_\xi\om^{IJ} \d\Si_{IJ}  - i_\xi \Si_{IJ}\w \d\om^{IJ}\right) = \int_{\p\Si} 2\pounds_\xi e_I\w\d e^I.
\ee
These have been studied in \cite{Godazgar:2020gqd} and previously in \cite{Corichi:2013zza}. The second equality follows from the identities \begin{subequations}\label{ixiS}\begin{align}
i_\xi\om^{IJ} \d\Si_{IJ} &= 2\pounds_\xi e_I\w\d e^I - 2d_\om\xi_I\w\d e^I, \label{ixiS1}\\
i_\xi\Si_{IJ}\w\d\om^{IJ} &= 2d(\xi_I\d e^I) - 2d_\om\xi_I\w\d e^I,
\end{align}\end{subequations}
and the fact that $\p\Si$ is compact.
Both identities \eqref{ixiS} can be proved starting from the torsionless condition and taking either variations or an inner product with $\xi$.
We use the term dual charges for short, but it should be kept in mind that we are talking about the contribution that the dual Lagrangian $\tl L^e$ gives to the usual charges. In particular, these dual contributions are associated with the usual diffeomorphism symmetries.

The dual charges are not affected by torsion, and therefore non-vanishing even in the absence of torsion. %
This is quite different from what happens in metric variables, where we have seen that zero torsion implies zero dual charges. It is useful to highlight the origin of the non-zero contribution. 
When torsion vanishes, the tetrad Holst Lagrangian is exactly zero, $\tl L^e = \Si_{IJ}\w F^{IJ}= T^I\w T_I - d(e_I\w T^I)=0$.
Accordingly, also its boundary variation is zero, $d\tl\th^e(\d)=0$. 
But the bare symplectic potential reduces to an exact 3-form, $\tl\th^e(\d)=\Si_{IJ}\w\d\om^{IJ}= -d(e_I\w\d e^I)$, instead of vanishing exactly like in the metric case.
Choosing to work with the bare symplectic potential means keeping the contribution to the phase space of this exact 3-form, and it has the effect of introducing dual charges absent in the metric formulation.
Hence, setting torsion to zero before or after extracting the bare symplectic potential from the Lagrangian leads to different phase spaces and different charges.\footnote{And it appears to be the reason why the existence of the dual charges was omitted in \cite{Frodden:2017qwh}. Taking instead a variation of the Nieh-Yan term $d(e_I\w T^I)$ gives a bare symplectic potential which is exactly zero in the absence of torsion, so no contribution to the dual charges in this case.} 
This shows the severe and even surprising consequences of picking a representative of a cohomology class instead of another.
Or, in other words, of working with different albeit field-equation-equivalent Lagrangians, a point recently emphasized in \cite{Freidel:2020xyx}.
The physical meaning of these choices is for us something that still needs to be clarified. For instance, a different viewpoint is that the freedom to add exact 3-forms, which was referred to as ambiguity of type $II$ of covariant phase space methods in \cite{Iyer:1994ys}, should not affect the  physical applications of the charges. As an example, the first law of black hole mechanics can be shown to hold independently of any choice for the symplectic potential \cite{Iyer:1994ys,DePaoli:2018erh}.

As a final remark about the dual charges, it is also clear from their expression \eqref{dualC} that they are not gauge-invariant (under the internal Lorentz transformations) unless $\d\Si^{IJ}$ vanishes, a property shared with the standard tetrad charges \eqref{HeS}. This fact is to be expected since $\pounds_\xi$ is not covariant under the internal Lorentz transformations, and can be dealt with in various ways. One is the Kosmann prescription discussed in Sec.~\ref{Sec:Kos}.

\section{Dressing the symplectic potential}
All the differences between Eq.~\eqref{Hg} and Eq.~\eqref{He} are to be traced back to Eq.~\eqref{thethg}.
If we choose 
\begin{equation}
\th'^{(e,\g)}(\d):=\th^{(e,\g)}(\d)+d\a(\d)
\end{equation}
as symplectic potential, we define a covariant phase space for tetrad gravity that matches exactly the metric one. 
In particular, we can see immediately that the ``dressed'' potential $\th'^{(e,\g)}$ reproduces all metric
Noether and Hamiltonian charges: 
\begin{align}
j'^{(e,\g)}(\d_\xi) &= \th'^{(e,\g)} - i_\xi L^{(e,\g)} 
= \th^{(g,\g)}(\d_\xi) - i_\xi L^{(g,\g)} = j^{(g,\g)}(\d_\xi), \\
\Om'^{(e,\g)}(\d,\d_\xi) &= \Om^{(e,\g)}(\d,\d_\xi) + \Om^\a(\d,\d_\xi) \nn\\
& = \int_\Si\d\th^{(e,\g)}(\d_\xi) -\d_\xi \th^{(e,\g)}(\d) - \th^{(e,\g)}([\d,\d_\xi]) + d\big[\d\a(\d_\xi) - \d_\xi\a(\d) -\a([\d,\d_\xi]) \big] \nn\\
& = \int_\Si\d\th^{(g,\g)}(\d_\xi) -\d_\xi \th^{(g,\g)}(\d) - \th^{(g,\g)}([\d,\d_\xi]) = \Om^{(g,\g)}(\d,\d_\xi).
\end{align}

It is actually instructive to see the restoration of the Hamiltonian charges explicitly.
The contribution of the DPS 2-form \eqref{alfa} to the generators 
\cite{Oliveri:2019gvm} can be conveniently written as 
\begin{align}\nn
\sd H_{(\l,\xi)}^{(\a,\g)} &  
= P_{IJKL} \int_{\p\Si} -\d(\Si^{IJ} e^{\r K})\pounds_\xi e_\r^L + \pounds_\xi ( \Si^{IJ} e^{\r K}) \d e^L_\r + \l^{IJ}\d \Si^{KL} \\
& = \int_{\p\Si} \d\k_\xi -\d q^e_\xi -i_\xi d\a_\star + \f12\eps_{IJKL} \l^{IJ} \d\Si^{KL} +\f1\g \int_{\p\Si}\l_{IJ} \d\Si^{IJ} -2\pounds_\xi e_I\w\d e^I 
\label{Ha}
\\\nn &=\sd H_{(\l,\xi)}^\a+\f1\g\sd \widetilde H_{(\l,\xi)}^\a,
\end{align}
where we used the identity
$    e^{\n[I}\pounds_\xi e^{J]}_\n = D^{[I}\xi^{J]}+i_\xi\om^{IJ}.
$
If we add Eq.~\eqref{Ha} to Eq.~\eqref{He}, the internal Lorentz charges cancel immediately, both standard and dual.
The cancellation of the dual gravitational charges is also immediate, thanks to the identities \eqref{ixiS}.\footnote{For compact $\p\Si$. If we were in a context where the boundaries of $\p\Si$ matter, one would need a further dressing of  the DPS 2-form by an exact 2-form to cancel this term.}
Finally, the matching of the \standard  diffeomorphism charges should be evident using Eq.~\eqref{HeS} and Eq.~\eqref{thethg}. Its explicit proof was presented in \cite{DePaoli:2018erh,Oliveri:2019gvm}.
For the current purpose of this note, what is important to stress is that the dual charges arise uniquely from the exact 3-form $d\a$ 
accounting for the discrepancy in Eq.~\eqref{thethg}.
It is therefore possible to set them to zero if one decides to use the ``dressed'' potential $\th'_{(e,\g)}$ to define the phase space of tetrad General Relativity.

\section{Isometries and the Kosmann derivative} \label{Sec:Kos}

A Killing vector Lie-drags the metric, but in general it will only Lie-drag the tetrad up to an internal Lorentz transformation. The corresponding notion of isometry for tetrad variables can be introduced using the Kosmann derivative (see e.g. \cite{TedMohd,Prabhu:2015vua}) 
\be\label{defK}
\mathcal{K}^{(e)}_{\xi}:=\pounds_{\xi} + \delta_{\bar{\lambda}}, \qquad \bar\l^{IJ}:=D^{[I}\xi^{J]}+i_\xi\om^{IJ}.
\ee
This transformation is defined for any diffeomorphism, and 
satisfies $\mathcal{K}^{(e)}_{\xi} e^I=0$ if $\xi$ is Killing. 
Notice that this transformation depends non-linearly on the tetrad itself, so it is a field-dependent gauge transformation.

Before 
the difference between symplectic potentials \eqref{thethg} was pointed out, 
it was proposed in  \cite{TedMohd,Prabhu:2015vua} 
to take as definition of charges in tetrad gravity those 
associated with the Kosmann derivative \eqref{defK}.
It is straightforward to see that 
\begin{equation}
\th^{(e,\g)}(\d_{\bar\l})\equiv d\a(\pounds_\xi)
\end{equation}
for any diffeomorphisms, for both standard and dual contributions of the potential
 \cite{Oliveri:2019gvm}; therefore we can understand the effectiveness of the Kosmann treatment from the more general relation \eqref{thethg}.
 In particular, the Noether charges of the Kosmann variations coincide with the metric charges for any diffeomorphism:
 \be
 j^{(e,\g)}(\mathcal{K}^{(e)}_{\xi}) = \th^{(e,\g)}(\pounds_\xi) +  \th^{(e,\g)}(\d_{\bar\l}) - i_\xi L^{(e,\g)}
 = \th^{(g,\g)}(\pounds_\xi)-  i_\xi L^{(g,\g)} =j^{(g,\g)}(\pounds_\xi),
 \ee
thanks to Eq.~\eqref{thethg}.  
We notice that the Noether charge computed in this way has no dual contribution from the Holst term.

For the Hamiltonian charges the situation is less straightforward, since they involve also (Lie derivatives of) $\th^{(e,\g)}(\d)$ for arbitrary variations. 
Let us see what the Kosmann prescription gives for the Hamiltonian charges. 
As shown in  \cite{Oliveri:2019gvm}, Eq.~\eqref{He} is valid also with a field-dependent parameter $\l^{IJ}$. Therefore, we can compute the charge associated with the Kosmann derivative using linearity of the generators and Eq.~\eqref{defK}, and obtain
\begin{align}
\Om^{(e,\g)}(\d,{\cal K}_\xi) &= \Om^{(e,\g)}(\d,{\pounds}_\xi)+\Om^{(e,\g)}(\d,\d_{\bar\l})\nn\\
&= \int_{\p\Si} P_{IJKL}\left(i_\xi\om^{IJ} \d \Si^{KL} - i_\xi\Si^{IJ}\w\d\om^{KL} -\bar\l^{IJ}\d\Si^{KL}\right) \label{match1}\\
\label{match}
&= \int_{\p\Si} -P_{IJKL}\delta\left( D^{[I}\xi^{J]} \Si^{KL}\right)- i_\xi \th^g + i_\xi d\a(\d) + P_{IJKL}  \Si^{IJ} \d \bar\l^{KL} \nn\\
&=\sd H^{g}_\xi + \int_{\p\Si}i_\xi d\a(\d) + P_{IJKL} \Si^{IJ} \d \bar\l^{KL}. 
\end{align}
Here we used that the first two terms of the third line reproduce the metric expression \eqref{Hg}, since the dual Komar 2-form $d\xi$ gives no contribution because $\p\Si$ is compact. For the same reason, we can replace $i_\xi d\a(\d)$ with $\pounds_\xi \a(\d)$. An explicit calculation,  reported in App.~\ref{AppA}, shows that 
 \begin{align}
\pounds_\xi \a(\d) &= -P_{IJKL}\Si^{IJ} \d\bar\l^{KL} +{\cal A}_\xi(\d), \label{Liea}\\
{\cal A}_\xi(\d)&:= \left[\f12\eps_{\m\n\r\s} (e^I_\a  \pounds_\xi g^{\a\r}+ \na_\a\xi^\a e^\r_I )
- \f1\g \pounds_\xi g_{\m\s} e_{\n}^I \right] \d e^{\s}_I dx^\m\w dx^\n.\label{OSform}
\end{align}
from this result, we conclude that the Hamiltonian charges associated to the Kosmann variation for an arbitrary diffeomorphism are
\be\label{OmA}
\Om^{(e,\g)}(\d,{\cal K}_\xi) = \sd H^{g}_\xi +\int_{\p\Si}{\cal A}_\xi(\d).
\ee
There are two remarks to make about this formula. First, in any situation where ${\cal A}_\xi$ does not vanish, the Kosmann prescription gives a different result than the charges in metric variables. 
Second, the whole dual contribution to any charge comes from the $\g$ contribution of the $\mathcal{A}_{\xi}(\d)$ term, marking the discrepancy between Kosmann and Lie variations.

Eq.~\eqref{OmA} allows us also to point out another important difference between standard and dual contributions. Consider the case of $\xi$ tangent to the 2d surface $\p\Si$. In this case, the standard contribution is manifestly integrable since the pull-back of $i_\xi\th^{(g,\g)}$ vanishes. The actual charge corresponds to the Noether charge, namely the Komar 2-form. The dual contribution is also integrable (this follows from similar considerations, but has also been shown explicitly in \cite{Godazgar:2020kqd}), but differs from the Noether charge, which vanishes.

We now specialize this formula to isometries and asymptotic symmetries, explaining how it allows us to recover different known results.

\subsection{Isometries}

Let us consider the case of $\xi$ being a Killing vector. Since $\pounds_{\xi} e^I = -\bar{\lambda}^{IJ}e_J$, an isometry gives non-zero dual charges \eqref{dualC}. These have the same structure of internal Lorentz charges, and are exactly cancelled by the $\d_{\bar\l}$ part of the Kosmann transformation. 
Therefore, the Kosmann prescription restores perfectly the expression \eqref{Hg} in metric variables for exact symmetries \cite{TedMohd,Prabhu:2015vua,Barnich:2016rwk,Frodden:2017qwh}. Both dual charges and internal Lorentz charges are absent.
From the perspective of Eq.~\eqref{OmA}, this result is established observing that $\mathcal{A}_\xi(\d)$ vanishes for isometries. Or equivalently,
\be
\pounds_\xi \a(\d) = -P_{IJKL}\Si^{IJ} \d\bar\l^{KL}
\ee
for $\xi$ Killing,
as shown already in \cite{DePaoli:2018erh}.

\subsection{Asymptotic symmetries at spatial infinity}

At spatial infinity, it is known that the expression \eqref{HeS} associated with the Lie derivative is enough to recover the Poincar\'e charges \cite{Ashtekar:2008jw}. Furthermore, there is no  contribution coming from $\tl L^e$ \cite{Corichi:2010ur}, so no dual contribution appears. These results can be understood because the whole contribution from $\a(\d)$ vanishes in this limit \cite{DePaoli:2018erh}: therefore the tetrad diffeomorphism charges match the metric ones. The same results can be reproduced using the Kosmann variation, because $\Om^{(e,\g)}(\d,\d_{\bar\l})$ also vanishes in this limit. 
This explains why it is possible to obtain the Poincar\'e charges from the Lie derivative, without explicitly studying the Kosmann variation. 
From the perspective of Eq.~\eqref{OmA}, the recovery of the Poincar\'e charges follows because the LHS reduces to $\Om^{(e,\g)}(\d,\d_\xi)$ in the limit, and ${\cal A}_\xi(\d)\mapsto 0$.

\subsection{Asymptotic symmetries at future null infinity} \label{AS}
%
 
What is more important and somewhat surprising, is that the Hamiltonian charges associated to the Kosmann variation,
 fail to reproduce the metric results for asymptotic symmetries at null infinity, albeit only in the sector of dual contributions. 
This is the main result of \cite{Godazgar:2020kqd}, and also the main motivation for our analysis. It can be immediately derived from Eq.~\eqref{OmA} and the
properties of the Bondi tetrad and BMS vectors used in
\cite{Godazgar:2020kqd}. 
With these properties, the Kosmann transformation preserves the boundary conditions at future null infinity \cite{Godazgar:2020kqd}.
 A simple calculation reported in App.~\ref{AppB} shows that  
\be \label{dualBMS2}
\int_{S^2_\infty}{\cal {A}}_\xi(\d) = \f1{2\g}\int_{S^2_\infty}  \d g_{BD} \left(\na_A\xi^D+\na^D\xi_A\right) dx^A\w dx^B. 
\ee
Here $A=2,3$ are sphere indices, but the whole spacetime metric is used to raise and lower indices. In particular, the standard piece (the first two terms of Eq.~\eqref{OSform}) vanishes, and the only  contribution comes from the dual piece (the third term of Eq.~\eqref{OSform}). The result coincides with 
the metric dual charges \eqref{dualC} computed in \cite{Godazgar:2018qpq,Godazgar:2018dvh}.
Therefore, it is the contribution of $\cal{A}_{\xi}$ in Eq.~\eqref{OmA} that is entirely responsible for the presence of dual charges at future null infinity $\scri$.

The careful reader may be puzzled by the equivalence between the dual charges \eqref{dualC} and the metric expression in \eqref{dualBMS2}.
In fact, we have explained that the dual charges arise from the difference between metric and tetrad bare symplectic potentials. This difference, given by \eqref{alfa}, is a quantity that cannot be written in metric form: it depends uniquely on the internal components of the tetrad.\footnote{If we define the tensorial tetrad perturbations $f_{\m\n}:=e_{I\n} \d e_\m^I$, it depends uniquely on the antisymmetric part, whereas it is the symmetric part alone that describes the metric perturbations.} Accordingly, the dual charges \eqref{dualC} depend also uniquely on the internal components of the tetrad. 
It is the fixing of coordinate and internal Lorentz gauges that makes it possible to turn these expressions into a pure metric form like \eqref{dualBMS2}.
This remarks highlights the important role of the gauge-fixing and working with a Bondi tetrad. See also the discussion on the use of adapted tetrads and time gauge in \cite{Oliveri:2019gvm}.

Let us briefly recall the derivation used in \cite{Godazgar:2020kqd}. 
The authors show that: $(i)$ the $\bar\l^{IJ}$ terms in Eq.~\eqref{match1}, both standard and dual, have vanishing limits at $\scri$, so that there is no contribution from  $\Om^{(e,\g)}(\d,\d_{\bar\l})$;
$(ii)$ the $\d\Si^{IJ}$ terms, both standard and dual, have vanishing limits; thus the only contribution comes from the $\d\om^{IJ}$ terms, that are gauge-invariant; $(iii)$ the limit of the $\d\om^{IJ}$ terms
reproduce the standard BMS terms, supplemented with the dual charges \eqref{dualC}, whose integrand can be rewritten as $2d_\om\xi_I\w\d e^I$ thanks to \eqref{ixiS1}. 
The last limits $(iii)$ can be also understood from the perspective of $\a(\d$). Using the given fall-off and gauge-fixing conditions, one finds that the standard term $\Om^{\a_\star}(\d,\d_\xi)$ vanishes, therefore the standard charge must coincide with the metric one. But the dual term $\Om^{\a_\g}(\d,\d_\xi)$ does not vanish, therefore the dual charges remain. 

\bigskip

Our discussion and slightly different derivation allow us to frame the results of \cite{Godazgar:2020kqd} in the broader context of the differences between tetrad and metric symplectic structures, and to derive the following lesson.
Using the Kosmann prescription,
dressing or not the symplectic potential is irrelevant for exact isometries or asymptotic symmetries at spatial infinity. But it makes a big difference for asymptotic symmetries at future null infinity: dual charges are either present or not, respectively without or with dressing. This can be summarized if we re-express 
Eq.~\eqref{OmA} as a difference between using the Kosmann prescription to compute charges, and adding the 2-form $\a(\d)$ to match the metric charges:
\begin{align}\label{29}
\Om^\a(\d,\d_\xi)  
&= \Om^{(e,\g)}(\d,\d_{\bar \l}) -\int_{\p\Si}{\cal A}_\xi(\d).
\end{align}

There are additional applications of this formula that can be anticipated.
While the Hamiltonian charges are usually most useful when associated to exact symmetries and asymptotic symmetries, there is a growing body of work in the literature concerned with charges beyond symmetries (e.g. \cite{Compere:2017wrj,Freidel:2020xyx}), and with extended BMS symmetries (e.g. \cite{Barnich:2009se,Campiglia:2014yka}). 
Eq.~\eqref{OmA} and Eq.~\eqref{29} can be used to study the effects of using tetrads in these more general contexts.

\section{Dual charges from cohomological methods}

The starting point for the cohomological methods is the field equations $\FE{e}$ and $\FE{\om}$ obtained varying the tetrad Lagrangian \eqref{Le}. Specializing the variation to a general gauge transformation, we have \cite{Hehl:1994ue,Barnich:2016rwk,Frodden:2017qwh,DeLorenzo:2018odq} 
\begin{align}\label{NdS}
\d_{(\lambda, \xi)} e^I\w \FE{e}_I + \d_{(\lambda, \xi)}\om^{IJ}\w \FE{\om}_{IJ}   = N(\lambda, \xi) + dS(\lambda, \xi),
\end{align}
with
\begin{align}
N(\l,\xi)&= (i_\xi\om^{IJ} - \l^{IJ}) (\FE{e}_I\w e_J - d_\om \FE{\om}_{IJ}) - i_\xi e^I d_\om \FE{e}_I +i_\xi T^I\w\FE{e}_I +i_\xi F^{IJ}\w\FE{\om}_{IJ}\nn \\
&= 2P_{IJKL} [(\l^{IJ}-i_\xi\om^{IJ}) e^K\w (d_\om T^L - F^{LM} \w e_{M}) + i_\xi e^I e^J\w d_{\om} F^{KL}],
\end{align}
and
\begin{align} \label{defS}
S(\l,\xi) &:= i_\xi e^I\FE{e}_I + (i_\xi\om^{IJ} - \l^{IJ})  \FE{\om}_{IJ} \nn\\
& = 2P_{IJKL}\left[ i_\xi e^I e^J\w F^{KL} + (i_\xi\om^{IJ} - \l^{IJ}) e^K\w T^L)\right].
\end{align}
The 3-form \eqref{defS} is closed on-shell thanks to Eq.~\eqref{NdS}, and it is taken as definition of Noether current in the cohomology approach. 
It differs from Wald's definition of Noether current \cite{Iyer:1994ys} by an exact form, which is nothing but $d\k_\xi$ in the metric case, and 
$d q^{e,\g}_{(\l,\xi)}$ in the tetrad case.
We notice that just like the Lagrangian defines the symplectic potential up to addition of an exact 3-form, also the field equations define $S(\l,\xi)$ up to addition of an exact 3-form.  
However the ``bare'' choice \eqref{defS} can be singled out as the unique current that is both closed and vanishing on-shell. 
For this reason, it is referred to as weakly-vanishing Noether current in the literature. This choice effectively eliminates the equivalent of ambiguity $II$ in the cohomological approach.

The BB charges are defined from the action of a certain homotopy operator $\mathcal{I}^{(3)}_{\d}$ on $S(\l,\xi)$, and are related to the Hamiltonian charges by
\be\label{BBcharge}
\sd Q^{\scr BB}_{(\l,\xi)}:=  \int_{\p\Si} \mathcal{I}^{(3)}_{\d}S(\l,\xi) \equiv \sd H^{e}_{(\l,\xi)} -  \int_{\p \Si}\f12 \mathcal{I}^{(3)}_{\d}\th(\d_{(\l,\xi)}),
\ee
see e.g.~\cite{Compere:2018aar,Oliveri:2019gvm}. For first order theories, $\th(\d)$ does not contain derivatives of the dynamical fields and thus lies in the kernel of the homotopy operator.
As a consequence, the BB charges always coincide with the Hamiltonian charges.
In fact, BB charges for the \standard Lagrangian of tetrad gravity were computed in  \cite{Barnich:2016rwk}, and shown to reproduce the standard Hamiltonian charges $\sd H^{e}_{(\l,\xi)}$ in Eq.~\eqref{He}. As for the dual charges, these can be easily deduced from the formulas of \cite{Barnich:2016rwk}, and were also explicitly given in \cite{Oliveri:2019gvm}, showing that they indeed reproduce the part 
$\sd \widetilde H^{e}_{(\l,\xi)}$ of Eq.~\eqref{He}.
Explicitly,
\be \label{QBB1st}
\sd Q^{BB}_{(\l,\xi)} = \int_\Si I^{(3)}_\d S(\l,\xi)
= \int_{\p\Si}P_{IJKL} \left[(i_\xi\om^{IJ}-\l^{IJ})\d \Si^{KL} - i_\xi\Si^{IJ}\w\d\om^{KL}\right].
\ee

To discuss the case of isometries or asymptotic symmetries, we can use again the Kosmann prescription. In fact, notice that Eq.~\eqref{QBB1st} is valid also with the field dependent parameter $\bar\l^{IJ}$ in place of $\l^{IJ}$, since $\bar\l^{IJ}$ does not contain derivatives of the fundamental fields and therefore lies in the kernel of $I^{(3)}_\d$.
The BB charges associated with the Kosmann variation are therefore given by the same calculation \eqref{match}, and we arrive at the same conclusions: the tetrad BB charges with the Kosmann variation reproduce the metric BB charges with the Lie variation in the case of exact symmetries (and recall that in this case, the BB metric charge coincides with Eq.~\eqref{Hg}). More in general, one obtains Eq.~\eqref{OmA}. For asymptotic symmetries, these recover the Poincar\'e charges at spatial infinity, and the BMS charges supplemented with the dual contribution at null infinity. 
We conclude that the dual contributions to the BMS charges can be computed also with cohomological methods, answering positively the question raised in the conclusions of \cite{Godazgar:2020kqd}.

This result puts the dual charges on firmer grounds, since the BB charges do not suffer from the same ambiguities of the Hamiltonian charges.\footnote{Even though the lack of ambiguity does not follow from additional physical input, but simply from the preferred choice of working with the unique weakly-vanishing Noether current. Let us add that if we give up this choice, we can add the exact 3-form $d\a(\d)$ to $S(\l,\xi)$, and the BB charges would then match the metric ones \cite{Oliveri:2019gvm}: no dual contributions in this case.
In this respect, let us point out that it is possible to use the homotopy operator to eliminate all ambiguities also in the symplectic potential,
by defining it as $\th(\d):={\cal I}^{(4)}_\d L$ \cite{anderson1992introduction,Barnich:2001jy}. This definition gives always the bare choice. It thus provides a precise mathematical framework in which the bare choice is preferred. 
This approach was recently emphasized in \cite{Freidel:2020xyx}.}
But it is not the end of the story, because on the other hand, the presence or not of the dual contribution depends on the order of the Lagrangian used.
If we start with the tetrad Lagrangian in second order form Eq.~\eqref{LeC}, with or without (con)torsion, the BB charges reproduce exactly the metric ones, see e.g. \cite{Oliveri:2019gvm}. 
This can be seen also directly from Eq.~\eqref{defS}: if one goes on-shell of the torsionless condition before or after applying the homotopy operator, the result differs in the dual sector (see also the analogue discussion at the end of Sec.~\eqref{dualorigin}).
Therefore the existence of the dual contribution depends crucially on the choice of first or second order formalism for the Lagrangian.

This leads us to a more general remark about cohomological methods. It is known that the charges they define are independent of the order of the Lagrangian (or more in general of field redefinitions \cite{Barnich:2000zw}) for exact symmetries (and for tetrad gravity, the notion of isometry is the Kosmann variation). For asymptotic symmetries, the analysis presented here shows that with the fall-off conditions considered, the independence of the order of the Lagrangian holds at spatial infinity but not at null infinity, albeit for the dual Lagrangian only.

\section{Conclusions} 
%

In this note, we drew attention to the fact that the dual contribution to the Hamiltonian charges arises from an exact 3-form in the tetrad symplectic potential originating from the Holst sector.
In the metric symplectic potential, on the contrary, the term in the symplectic potential originating from the Holst sector vanishes exaclty in the absence of torsion. 

A new technical result of this work is the computation of the (quasi-local) Hamiltonian charges associated to general diffeomorphisms via the Kosmann derivative \eqref{OmA}. 
This formula allows us to give a comprehensive understanding of various results in the literature, and to expose the presence of discrepancies with the metric results when the diffeomorphism is not a Killing vector. 

An interesting open question at this point is whether  
there exist slightly different fall-off conditions to future null infinity, such that the Kosmann prescription matches exactly the metric charges, including vanishing dual contributions, as it does at spatial infinity. 
A similar question for future work is whether 
the discrepancy ${\cal A}_\xi(\d)$ persists at spatial infinity under the fall-off conditions considered in 
\cite{Henneaux:2020ekh} to enlarge the Poincar\'e symmetry group.

We also pointed out that using the ambiguity of covariant phase space methods, it is possible to redefine the tetrad phase space so that the dual contributions are exactly zero. There is also a ``half-dressing'' choice, adding only $d\a_\star(\d)$ and not $d\a_\g(\d)$, that allows one to keep the dual contributions, while being guaranteed to recover the standard metric results for arbitrary diffeomorphisms.
Ultimately, one needs a physical prescription to fix the freedom of making these choices. For instance, the possibility of dressing the symplectic potential adding exact 3-forms has been advocated to extend the asymptotic symmetries to include superrotations \cite{Compere:2018ylh}. If a dressing is added, it would also be useful to have an interpretation for it in terms of a boundary Lagrangian depending on new fields, as done for instance in  \cite{Wieland:2019hkz,Freidel:2020xyx,Freidel:2020svx}. From this point of view, we recall that the internal Lorentz charges are set identically to zero with the boundary Lagrangian of \cite{Wieland:2019hkz}. It would be interesting to see what happens to the dual contributions to the charges in this set-up.

In spite of the derivation of dual contributions to the gravitational charges via cohomological methods, 
their presence at future null infinity appears to us still on a more ambiguous footing than the standard BMS charges. In fact, the  
former depend on the variables used and on the order of the Lagrangian, unlike the latter.
It would then be preferable to have a physical effect that could be directly linked to these contributions to settle the issue.
Some of the dual contributions are related to the NUT charges and thus absolutely conserved  \cite{ashtekar1982nut}. 
But a crucial question in this respect is whether there are dual contributions that contain radiative modes, and would impact the flux-balance laws of the standard BMS charges (see e.g. \cite{Compere:2019gft}).

\subsection*{Acknowledgments}
The authors thank Abhay Ashtekar for helpful discussions, and Glenn Barnich, Uri Kol and Ali Seraj for uselful correspondence. We are especially grateful to Mahdi Godazgar for extensive discussions about subleading terms which led us to sharpen our results.
R.O. is funded by the European Structural and Investment Funds (ESIF) and the Czech Ministry of Education, Youth and Sports (MSMT), Project CoGraDS - CZ.02.1.01/0.0/0.0/15003/0000437. 


\begin{appendix}
\setcounter{equation}{0}
\renewcommand{\theequation}{\Alph{section}.\arabic{equation}}

\section{The Lie vs. Kosmann discrepancy} \label{AppA} 
In this Appendix we report the calculation of Eq.~\ref{Liea}.
As discussed in the main text, a way to compare the Kosmann Hamiltonian charges with the tetrad potential and the Lie Hamiltonian charges with the metric potential, is to see whether the Kosmann contribution $\th^{(e,\g)}(\d_{\bar\l})$ matches the Lie derivative of the DPS 2-form \eqref{alfa}. 
To that end, we compute  $\pounds_{\xi}\a_{\star}$ and $\pounds_{\xi}\a_{\g}$. The latter is a straightforward calculation: 
\be
\pounds_\xi (e_{I[\mu}\d e^I_{\n]}) = -\pounds_{\xi}g_{\s[\m} e^I_{\n]}\d e^{\s}_I + (\pounds_{\xi}e^I_{[\mu})\d e_{\n]I} + e_{I[\mu} \pounds_{\xi} \d e^{I}_{\nu]} = -\pounds_{\xi}g_{\s[\m} e^I_{\n]}\d e^{\s}_I - \Si_{IJ}\d \bar{\l}^{IJ}.
\ee
The computation of $\pounds_{\xi}\a_{\star}$ is more involved. We first notice the identity
\begin{align}
\eps_{\m\r\a\b} &e^\a_I e^\b_J \pounds_\xi e_\n^I \d e^{\r J} = \eps_{\m\n\a\b} e^\a_I e^\b_J \pounds_\xi e_\r^I \d e^{\r J} 
+\f12 \eps_{\m\l\a\b}\eps^{\l\s\g\d}\eps_{\g\d\n\r} e^\a_I e^\b_J \pounds_\xi e_\s^I \d e^{\r J} \nn\\
&=\eps_{\m\n\a\b} e^\a_I e^\b_J \pounds_\xi e_\r^I \d e^{\r J} 
-\eps_{\m\r\a\b} e^\a_I e^\b_J \pounds_\xi e_\n^I \d e^{\r J} + \eps_{\m\n\b\r}  \na_\a\xi^\a e^\b_I \d e^{\r I}  - \eps_{\m\a\n\r} \pounds_\xi e^\a_I \d e^{\r I}, 
\end{align}
from which we have 
\be\label{pruzzo}
\eps_{\m\r\a\b} e^\a_I e^\b_J \pounds_\xi e_\n^I \d e^{\r J} =\f12 \eps_{\m\n\a\b} e^\a_I e^\b_J \pounds_\xi e_\r^I \d e^{\r J}  
+ \f12 \eps_{\m\n\b\r} \na_\a\xi^\a e^\b_I \d e^{\r I}+ \f12 \eps_{\m\n\a\r} \pounds_\xi e^\a_I \d e^{\r I}.
\ee
Then, we compute 
\begin{align}
& \f12\eps_{IJKL}\Big[\Si^{IJ}\d e^{\r K}\pounds_\xi e_\r^L -\pounds_\xi(\Si^{IJ}e^{\r K})\d e_\r^L\Big] \nn\\& 
\qquad = \f12 \eps_{\m\n\a\b} e^\a_I e^\b_J \d e^{\r I} \pounds_\xi e_\r^J + \f12 \eps_{\m\n\a\r} e^\a_J e_{\b I} \d e^{\r I} \pounds_\xi e^{\b J}
 + \eps_{\m\r\a\b} e^\a_I e^\b_J \pounds_\xi e_\n^I \d e^{\r J} \nn\\ &\qquad 
= -\f12 \eps_{\m\n\a\b} e^\a_I e^\b_J  \pounds_\xi e_\r^I \d e^{\r J} + \f12\eps_{\m\n\a\r} e^I_\b \d e^\r_I \pounds_\xi g^{\a\b}
-\f12 \eps_{\m\n\a\b} \pounds_\xi e^\a_I \d e^{\b I}  + \eps_{\m\r\a\b} e^\a_I e^\b_J \pounds_\xi e_\n^I \d e^{\r J}\nn\\
&\qquad = \f12\eps_{\m\n\a\r} e^I_\b \d e^\r_I \pounds_\xi g^{\a\b}+ \f12 \eps_{\m\n\b\r} \na_\a\xi^\a e^\b_I \d e^{\r I}\nn\\
&\qquad = \f12\eps_{\m\n\r\s} \left( e^I_\a  \pounds_\xi g^{\a\r} + \na_\a\xi^\a e^\r_I \right) \d e^\s_I.
\end{align}
In the second equality we exchanged $I$ and $J$ in the first term and expanded $e^\a_J \pounds_{\xi}e^{\b J}$ in the second term. In the third equality we used \ref{pruzzo}.
This quantity vanishes for a Killing vector, and measures the failing of the Kosmann prescription to reproduce the metric Hamiltonian generator for a non-Killing diffeomorphism.

\section{Limit at future null infinity}\label{AppB}

We use Bondi coordinates $x^\m=(u,r,x^A)$, with $\{A, B, \dots\}$ denote coordinates indices on the 2-sphere.
The tetrad used in \cite{Godazgar:2020kqd} is chosen in doubly-null form, with $e^0_\m e^{1\m}=-1$ and $e^2_\m e^{3\m}=1$, and adapted to the Bondi coordinates, as follows: 
\begin{subequations}
\begin{align}
    e^0_{\mu}dx^{\mu} &= \f12 F du+dr,  \quad &e^{\mu}_0\p_\mu &= \p_r,\\
    e^1_{\mu}dx^\m &=e^{2\beta}du, \quad &e^{\m}_{1}\p_\m&=e^{-2\beta}\left(\p_u-\f12 F\p_r +C^{A}\p_A\right),\\
    e^i_\m dx^\m &=r E^i_A\left(dx^A-C^Adu \right), \quad &e^\m_{i}\p_\m &=\f1r E^A_i \p_A,
\end{align}
\end{subequations}
where $i=2,3$, and $(E^i_A, E_i^A)$ are the dyad and its inverse on the 2-sphere. The fall-off behaviour of the functions $F$, $\beta$, $E^i_A$ and $C^A$ can be found in \cite{Godazgar:2020kqd}. 
We now evaluate ${\cal A}_\xi(\d)$ in these coordinates.

The standard contribution, pulled back on the 2-sphere, gives
\begin{align} \label{Astandard}
  \f12\eps_{\m\n\r\s} &\left(e^I_\a  \pounds_\xi g^{\a\r}+ \na_\a\xi^\a e^{\r I}\right)\d e^{\s}_I
  \nn\\ &\stackrel{S^2}{=}\f12\eps_{A B u r} \left[e^1_u \d e^{r}_1 \pounds_\xi g^{uu}  -e^1_u\d e^{u}_1 \pounds_\xi g^{ur} +\nabla_\a \xi^\a \left(e^{u1}\d e^{r}_1 -e^{r1}\d e^{u}_1 \right)\right] \nn\\
  &=\f12\eps_{A B u r}\left[\left(F \d\b -\f12 \d F\right) \pounds_\xi g^{uu}  +2\d\b\left( \pounds_\xi g^{ur} -e^{-2\b}\nabla_\a \xi^\a \right)\right].
\end{align}
Using the familiar fall-off conditions at future null infinity, we have $\pounds_\xi g^{uu} = 0$, whereas $\pounds_\xi g^{ur}= o(r^{-1})$ and $\nabla_\a \xi^\a = o(r^{-1})$. This leaves a priori the possibility of a non-vanishing contribution at subleading order. 
However, the BMS vector fields satisfy the non-trivial properties \cite{Godazgar:2020kqd}:
$\nabla_r \xi^{u}=0$, $g_{\m (r}\nabla_{A)} \xi^\mu =0$, and $\nabla_A \xi^A = C^A \nabla_A \xi^u$,
which imply that $\pounds_\xi g^{ur} =e^{-2\b}\nabla_\a \xi^\a$.
These properties ensue that \eqref{Astandard} vanishes exactly. One may worry that this vanishing is coordinate-dependent, but there is at least one simple test of minimal coordinate-independence that can be done. In the Newman-Unti gauge, one chooses the same coordinates except for $r$, which is taken to be an affine parameter along the null rays, instead of the radius of the spheres. With this choice, $\b\equiv 0$ and the vanishing of  \eqref{Astandard} is straightforward.
The covariant form of ${\cal A}_\xi$ shows the properties that $\xi$ must satisfy in order to have a completely coordinate-invariant statement.

The dual contribution gives 
\begin{align}
    \f1\g \pounds_\xi g_{\s[\m} e^{\s}_I \d e_{\n]}^I &\stackrel{S^2}{=} 
    \f1\g\left( \nabla_{[A} \xi_{\s} + \nabla_{\s} \xi_{[A} \right) e^{\s}_i \d e_{B]}^i \nn\\
    &=\f1\g\left( \nabla_{[A} \xi_{C} + \nabla_{C} \xi_{[A} \right) e^{C}_i \d e_{B]}^i  \nn\\
    &=\f1{2\g}\left( \nabla_{[A} \xi_{C} + \nabla_{C} \xi_{[A} \right) g^{CD} \d g_{B]D } 
    \nn\\
    &=\f1{2\g}\left[\left( \nabla_{[A} \xi^{D} + \nabla^{D} \xi_{[A} \right) -g^{rD} \left(\nabla_{[A} \xi_r + \nabla_r \xi_{[A}  \right)\right]\d g_{B]D } 
    \nn\\
    &=\f1{2\g}\left( \nabla_{[A} \xi^{D} + \nabla^{D} \xi_{[A} \right)\d g_{B]D }.
\end{align}
In this derivation we used crucially various properties that hold thanks to the use of Bondi coordinates. In particular, $g^{uA} =0=g_{rB}$, $g_{BD} =e^i_B e_{Di} =r^2 E^i_B E_{Di}$ with $g^{CD}=r^{-2}E^{Ci}E^{D}_i$ its inverse. And the properties of the BMS vector fields $\xi^\m$ used above. 
This finishes the proof of Eq.~\eqref{dualBMS2}.

\end{appendix}


\bibliographystyle{JHEPs}
\bibliography{bibliosimo}

\end{document}